\documentclass[12pt]{article}

\input tcilatex
\QQQ{Language}{
American English
}

\begin{document}

\author{I. Radinschi* and I-Ching Yang$^{**}$ \\
Department of Physics \\
``Gh. Asachi'' Technical University, \\
Iasi, 6600, Romania and\\
Department of Natural Science Education, \\
National Taitung Teachers College, Taitung, \\
Taiwan 950, Republic of China \\
*iradinsc@phys.tuiasi.ro, $^{**}$icyang@dirac.phys.ncku.edu.tw}
\title{On the M\o ller Energy-Momentum Complex of the Melvin Magnetic Universe}
\date{}
\maketitle

\begin{abstract}
We use the M\o ller energy-momentum complex to calculate the energy of the
Melvin magnetic universe. The energy distribution depends on the magnetic
field.


PACS: 04. 20.-q, 04. 20.
\end{abstract}

\section*{Introduction}

The subject of energy-momentum localization in general relativity continues
to be an open one because there is no given yet a generally accepted
expression for the energy-momentum density. Even they are coordinate
dependent, various energy-momentum complexes give the same energy
distribution for a given space-time.

Aguirregabiria, Chamorro and Virbhadra [1] obtained that the energy-momentum
complexes of Einstein [2], Landau and Lifshitz [3], Papapetrou [4], and
Weinberg [5] give the same result for the energy distribution for any
Kerr--Schild metric. Also, recently, Virbhadra investigated [6] if these
definitions lead to the same result for the most general nonstatic
spherically symmetric metric and found they disagree. Only the
energy-momentum complex of Einstein gives the same expression for the energy
when the calculations are performed in the Kerr--Schild Cartesian and
Schwarzschild Cartesian coordinates. The M\o ller energy-momentum complex
allows to compute the energy in any coordinate system.

Some results recently obtained [7]-[10] sustain that the M\o ller
energy-momentum complex is a good tool for obtaining the energy distribution
in a given space-time. Also, in his recent paper, Lessner [11] gave his
opinion that the M\o ller definition is a powerful concept of energy and
momentum in general relativity. Very important is the Cooperstock [12]
hypothesis which states that the energy and momentum are confined to the
regions of non-vanishing energy-momentum tensor of the matter and all
non-gravitational fields. Also, Chang, Nester and Chen [13] showed that the
energy-momentum complexes are actually quasilocal and legitimate expression
for the energy-momentum.

In this paper we calculate the energy distribution of the Melvin magnetic
universe in the M\o ller prescription. We use geometrized units ($G=1, c=1$)
and follow the convention that Latin indices run from 0 to 3.

\section*{Energy in the M\o ller Prescription}

The Melvin magnetic universe [14], [15] is described by the electrovac
solution to the Einstein--Maxwell equations and consists in a collection of
parallel magnetic lines of forces in equilibrium under their mutual
gravitational attraction. The Einstein--Maxwell equations are

\begin{equation}
R_i^{\;k}-{\frac 12\,}g_i^{\;k}\,R=8\,\pi \,T_i^{\;k},
\end{equation}
\begin{equation}
{\frac 1{\sqrt{-g}}}\left( \sqrt{-g}\,F^{ik}\right) _{,k}=4\,\pi \,J^i,
\end{equation}
\begin{equation}
F_{ij,k}+F_{jk,i}+F_{ki,j}=0.
\end{equation}

The energy-momentum tensor of the electromagnetic field is given by

\begin{equation}
T_i^{\;k}={\frac 1{4\,\pi }}\left[ -F_{im}\,F^{km}+{\frac 14\,}%
g_i^{\;k}\,F_{mn}\,F^{mn}\right] .
\end{equation}

The electrovac solution corresponds to $J^i=0$ and is given by the metric

\begin{equation}
ds^2=L^2\,(dt^2-dr^2-r^2\,d\theta ^2)-L^{-2}\,\,r^2\,\sin ^2\theta
\,\,d\varphi ^2,  \label{5}
\end{equation}
where

\begin{equation}
L=1+{\frac 14\,}B_0^2\,r^2\,\sin ^2\theta .  \label{6}
\end{equation}

The Cartan components of the magnetic field are

\begin{equation}
\begin{array}{l}
H_r=L^{-2}\,\,B_0\,\cos \theta , \\ 
H_\theta =-L^{-2}\,\,B_0\,\sin \theta .
\end{array}
\end{equation}
$B_0$ is the magnetic field parameter and is a constant in the solution
given by (\ref{5}) and (\ref{6}).

The energy-momentum tensor has the non-vanishing components

\begin{equation}
\begin{array}{l}
\displaystyle T_1^{\;1}=-T_2^{\;2}={\frac{B_0^2\,(1-2\,\sin ^2\theta )}{%
8\,\pi \,L^4}}, \\ 
\displaystyle T_0^{\;0}=-T_3^{\;3}={\frac{B_0^2}{8\,\pi \,L^4}}, \\ 
\displaystyle T_2^{\;1}=-T_1^{\;2}={\frac{2\,B_0^2\,\sin \theta \,\cos
\theta }{8\,\pi \,L^4}}.
\end{array}
\end{equation}

The M\o ller energy-momentum complex $M_i^{\;k}$ [16] is given by

\begin{equation}
M_i^{\;k}={\frac 1{8\,\pi }\,}\chi _{i\;\;\;,l}^{\;kl},
\end{equation}
where

\begin{equation}
\chi _i^{\;kl}=-\chi _i^{\;lk}=\sqrt{-g}\,\left( {\frac{\partial g_{in}}{%
\partial x^m}}-{\frac{\partial g_{im}}{\partial x^n}}\right)
\,g^{km}\,g^{nl}.
\end{equation}

Also, $M_i^{\;k}$ satisfies the local conservations laws

\begin{equation}
{\frac{\partial M_i^{\;k}}{\partial x^k}}=0.
\end{equation}

$M_0^{\;\,0}$ is the energy density and $M_\alpha ^{\;\,0}$ are the momentum
density components.

The energy and momentum are given by

\begin{equation}
E=\int \hskip-7pt\int \hskip-7pt\int M_0^{\,\,\,0}dx^1dx^2dx^3={\frac
1{8\,\pi }}\int\hskip-7pt\int\hskip-7pt\int {\frac{\partial \chi _0^{\,\,0l}%
}{\partial x^l}\,}dx^1\,dx^2\,dx^3.
\end{equation}

For the Melvin magnetic universe we obtain

\begin{equation}
\begin{array}{l}
\displaystyle \chi _0^{\;01}={\frac{\,B_0^2\,r^3\,\sin ^3\theta }{%
(1+1/4\,B_0^2\,r^2\,\sin ^2\theta )}}, \\ 
\displaystyle \chi _0^{\;02}={\frac{\,B_0^2\,r^2\,\cos \theta \,\sin
^2\theta }{(1+1/4\,B_0^2\,r^2\,\sin ^2\theta )}}.
\end{array}
\end{equation}

After some calculations, applying the Gauss theorem and plugging (13) into
(12) we obtain the energy distribution

\begin{equation}
E(r)={\frac 13\,}B_0^2\,r^3-{\frac 1{15}\,}B_0^4\,r^5+{\frac 1{70}\,}%
B_0^6\,r^7.
\end{equation}

Put the $G$ and $c$ at their places we get

\begin{equation}
E(r)={\frac 13\,}B_0^2\,r^3-{\frac 1{15}\,\frac G{c^4}\,}B_0^4\,r^5+{\frac
1{70}\,\frac{G^2}{c^8}\,}B_0^6\,r^7.
\end{equation}

The first term represents twice of the classical value of energy [17]
obtained in the Landau and Lifshitz and Papapetrou prescriptions. The other
terms are due to the relativistic correction.

\section*{Discussion}

Many results recently obtained sustain the viewpoint of Bondi [18]. He gave
his opinion that a nonlocalizable form of energy is not admissible in
relativity.

We obtain the energy distribution of the Melvin magnetic universe using the
energy-momentum complex of M\o ller. The energy depends on the magnetic
field. The result is different as that obtained by Xulu [17] using the
energy-momentum complexes of Landau and Lifshitz and those of Papapetrou.
The first term represents twice of the classical value of energy [17]
obtained in the Landau and Lifshitz and Papapetrou prescriptions. The third
term, that is due to the relativistic correction, is twice of value of
energy obtained in [17]. Also, the M\o ller energy-momentum complex does not
need to carry out calculations in any particular coordinates.

\section{Acknowledgments}

I. -C. Yang thanks the National Science Council of the Republic of China for
financial support under the contract number NSC 90-2112-M-143-003.


\begin{thebibliography}{99}
\bibitem{1}  J. M. Aguirregabiria, A. Chamorro and K. S. Virbhadra, Gen.
Rel. Grav. \textbf{25\ }, 1393 (1996).

\bibitem{2}  C. M\o ller, Ann. Phys. (N.Y.) \textbf{4}, 347 (1958); A.
Trautman, in \textit{Gravitation: an Introduction to Current Research}, ed.
L. Witten (Wiley, New York, 1962, p.169); R.P. Wallner, Acta Physica
Austriaca \textbf{52}, 121 (1980).

\bibitem{3}  L.D. Landau and E.M. Lifshitz, \textit{The Classical Theory of
Fields} (Pergamon Press, 1987, p.280).

\bibitem{4}  A. Papapetrou, Proc. R. Irish. Acad. \textbf{A52}, 11 (1948).

\bibitem{5}  S. Weinberg, \textit{Gravitation and Cosmology: Principles and
Applications of General Theory of Relativity} (John Wiley and Sons, New
York,1972, p.165).

\bibitem{6}  K. S. Virbhadra, Phys. Rev\textit{.} \textbf{D60}, 104041
(1999).

\bibitem{7}  I-Ching Yang, Wei-Fui Lin and Rue-Ron Hsu, Chin. J. Phys. 
\textbf{37}, 113 (1999).

\bibitem{8}  S.S. Xulu, gr-qc/0010062
\bibitem{9}  S.S. Xulu, gr-qc/0010068

\bibitem{10}  I. Radinschi, Mod. Phys. Lett. \textbf{A16(10)}, 67 (2001).

\bibitem{11}  G. Lessner, Gen. Relativ. Gravit. \textbf{28}, 527 (1996).

\bibitem{12}  F.I. Cooperstock, Mod. Phys. Lett. \textbf{A14}, 1531 (1999).

\bibitem{13}  Chia-Chen Chang, J. M. Nester and Chiang-Mei Chen, Phys. Rev.
Lett. \textbf{83}, 1897 (1999).

\bibitem{14}  M.A. Melvin, Phys. Lett. \textbf{8}, 65 (1964).

\bibitem{15}  F.J. Ernst, J. Math. Phys. \textbf{15}, 1409 (1974).

\bibitem{16}  C. M\o ller, Ann. Phys. (N.Y.) \textbf{4}, 347 (1958).

\bibitem{17}  S. S. Xulu, Int. J. Mod. Phys. \textbf{A15}, 4849 (2000).

\bibitem{18}  H. Bondi, Proc. R. Soc. London \textbf{A427}, 249 (1990).
\end{thebibliography}
\end{document}